\documentclass{cpbtex}

\usepackage{amsmath,graphicx,amssymb,subfigure}
\usepackage{slashed}
\usepackage{subfigure}
\usepackage{isomath}
\usepackage{enumerate}
\usepackage{stmaryrd}
\usepackage{amsfonts}
\usepackage{cite}
\def\be{\begin{equation}}
\def\ee{\end{equation}}
\def\beq{\begin{equation}}
\def\eeq{\end{equation}}
\def\ba{\begin{array}}
\def\ea{\end{array}}
\newcommand{\bea}{\begin{eqnarray}}
\newcommand{\eea}{\end{eqnarray}}

\begin{document}
\begin{CJK*}{GBK}{song}

\title{Holographic Heat Engine Efficiency of Hyperbolic Charged Black Holes\thanks{This work is partly supported by NSFC (No.11875184).}}


\author{Wei Sun, \ Xian-Hui Ge \thanks{Corresponding author. E-mail:gexh@shu.edu.cn}\\
Department of Physics, Shanghai University, Shanghai 200444, China\\  
}

\maketitle

\begin{abstract}
We consider a four-dimensional charged hyperbolic black hole as  working matter to establish a black hole holographic heat engine, and use the rectangular cycle to obtain the heat engine efficiency.
We find that when the increasing of entropy is zero, the heat engine efficiency of the hyperbolic black hole becomes the well-known Carnot efficiency. We also find that less charge corresponds to higher efficiency in the case of $\tilde{q}>0$. Furthermore, we study the efficiency of the flat case and  spherical case and compare the efficiency with that of the hyperbolic charged black holes. Finally, we use numerical simulation to study the efficiency in benchmark scheme.
\end{abstract}

\textbf{Keywords:} Holographic heat engines, Benchmark cycle, Hyperbolic charged black holes

\textbf{PACS:}  95.30.Tg; 04.70.-s; 04.70.Dy

\section{Introduction}
\begin{figure}
\centering
\includegraphics[width=5cm]{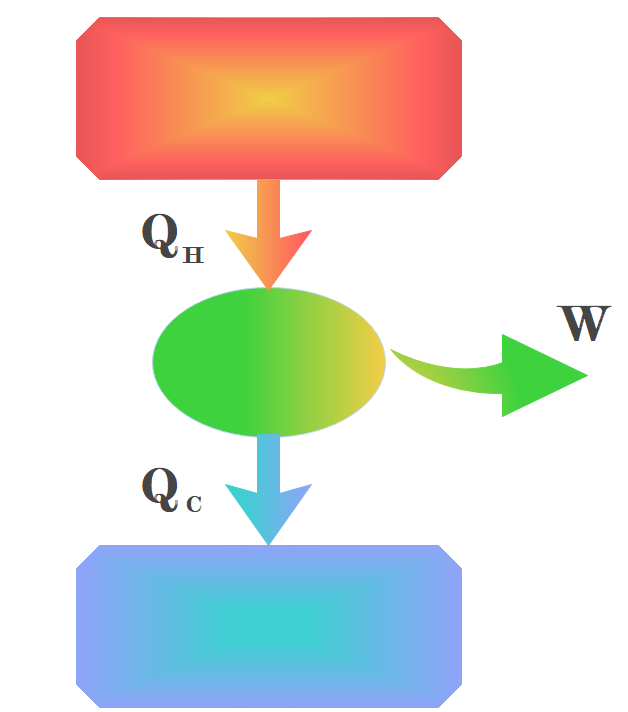}
\caption{The usual heat engine.}
\label{Heat engine schematic}
\end{figure}

\begin{figure}
\centering
\includegraphics[width=6cm]{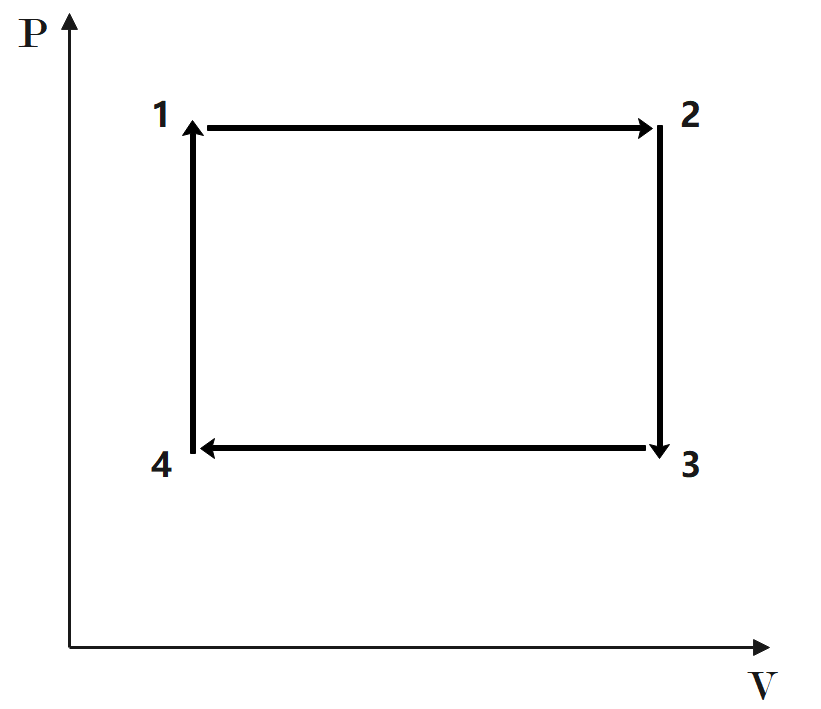}
\caption{Rectangular cycle in P-V diagram.}
\label{P-V diagram}
\end{figure}

Recently, some authors consider the cosmological constant as the pressure in the thermodynamics of black holes\cite{Kastor:2009wy,Kastor:2010gq,Kubiznak:2012wp,Kubiznak:2014zwa}, which has led to many interesting related studies. For example, some interesting black hole thermodynamic phenomena and rich phase structures similar to the van der Waals fluids are discovered\cite{Kubiznak:2016qmn}. Other authors have built a bridge between extended black hole thermodynamics and quantum complexity\cite{Sun:2019yps}. In 2014, Johnson proposed a concept of a holographic heat engine by considering AdS black holes as working matter\cite{Johnson:2014yja}. We show a schematic diagram of the usual heat engine in Fig.\ref{Heat engine schematic}.
In general, the heat engine efficiency can be expressed as
\be
\eta=\frac{W}{Q_{H}}=1-\frac{Q_{C}}{Q_{H}},
\ee
where  $Q_{H}$ and $Q_{C}$ are the heat from the high-temperature source to the heat engine and the heat from the heat engine to the low-temperature source, respectively.
The efficiency of the heat engine depends on the path selection on the P-V diagram. For the Carnot cycle, the path consists of two isotherms and two adiabatic lines on the P-V diagram, and the change in entropy is zero. The efficiency of the Carnot cycle with temperature $T_{H}$ and $T_{C}$ is $\eta_{C}=1-T_{C} / T_{H}$, which is the maximum efficiency that can be achieved. However, the P-V diagram of the AdS black hole is complicated, and difficult to establish the Carnot cycle. Johnson proposed a rectangular cycle, which consists of two equal-pressure processes and two equal-volume processes(Fig.\ref{P-V diagram}).
\
The rectangular cycle is equivalent to the Carnot cycle of a static black hole, because in this case, both entropy and volume depend on the single variable of the horizon radius $r_{+}$. In other words, we can replace the volume on the horizontal axis in Fig.\ref{P-V diagram} in entropy. In this case, the specific heat capacity of a constant volume is zero $C_{V}=0$, the equal-volume process is the same as the adiabatic process. The specific heat capacity of a constant pressure $C_{p}$ can be calculated by the radius of the horizon.
 $Q_{H}$ and $Q_{C}$ can be expressed as\cite{Johnson:2016pfa}
\begin{equation}
Q_{H}=\int_{T_{1}}^{T_{2}} C_{P} d T=\int_{S_{1}}^{S_{2}} T\left(\frac{\partial S}{\partial T}\right)\left(\frac{\partial T}{\partial S}\right) d S=\int_{S_{1}}^{S_{2}} T d S,
\end{equation}\label{1}
\begin{equation}
Q_{c}=\int_{T_{4}}^{T_{3}} C_{P} d T=\int_{S_{4}}^{S_{3}} T\left(\frac{\partial S}{\partial T}\right)\left(\frac{\partial T}{\partial S}\right) d S=\int_{S_{4}}^{S_{3}} T d S.
\end{equation}\label{2}
The numbers in the subscript represent the four stages in the cycle shown in Fig.\ref{P-V diagram}. The heat engine efficiency is
\begin{equation}
\eta=1-\frac{Q_{C}}{Q_{H}}=1-\frac{\int_{S_{3}}^{S_{4}} T d S}{\int_{S_{1}}^{S_{2}} T d S}.
\end{equation}
In  the AdS black hole thermodynamics, mass is defined as enthalpy $\mathrm{d}H=T \mathrm{d}S+V \mathrm{d}P$. In the rectangular cycle, the pressure remains constant during the heat exchange process i.e., $dP=0$. Combined $Q_{H}$ with $Q_{c}$ , the heat exchange can be regarded as the change of the black hole mass M,
\begin{equation}
Q_{H}=\int_{S_{1}}^{S_{2}} T d S=M_{2}-M_{1}, \quad Q_{c}=\int_{S_{4}}^{S_{3}} T d S=M_{3}-M_{4},
\end{equation}
where $M_{1},M_{2},M_{3},M_{4}$ respectively correspond to the black hole mass in the four stages of the cycle.  Johnson proposed an exact efficiency formula for the holographic black hole heat heat engines\cite{Johnson:2016pfa}
\begin{equation}\label{exact efficiency}
\eta_{f}=1-\frac{M_{3}-M_{4}}{M_{2}-M_{1}}.
\end{equation}

The efficiency of black hole heat engine is an important part of the study of black hole thermodynamics. However, most authors only discuss the case with the horizon being a positive constant curvature hypersurface\cite{Jafarzade:2017sur,Setare:2015xaa,Johnson:2015fva,Zhang:2016wek,Johnson:2015ekr,Wei:2017vqs,Mo:2017nes,Hendi:2017bys,Xu:2017ahm,Mo:2017nhw,Hennigar:2017apu,Zhang:2018vqs,21,22}. Note that there are huge differences in the thermal stability of black holes in AdS spaces with different curvatures. It is also necessary to study the heat engine efficiency of the AdS black holes with negative constant curvature horizon. In fact, some authors have studied the thermal stability of hyperbolic black holes\cite{Cai:2004pz}. Based on the study of thermodynamics of hyperbolic black holes, in this paper we further discuss the efficiency of the holographic heat engine of hyperbolic black holes. In the next section we briefly review its thermodynamics, then consider charged hyperbolic black holes as working matter to establish a rectangular heat engine cycle, and further discuss the effect of electric charge and the efficiency of the flat case and spherical case. In section 3, we calculate the efficiency of benchmark cycle by using numerical simulation, and make a summary and discussion in the last section.

\section{Hyperbolic charged black hole as a heat engine}
The metric of the $(n + 2)$-dimensional hyperbolic charged AdS black hole is
\be
d s^{2}=-f(r) d t^{2}+f(r)^{-1} d r^{2}+r^{2} d \Sigma_{n}^{2}\label{metric},
\ee
where
\be
f(r)=k-\frac{m}{r^{n-1}}+\frac{\tilde{q}^{2}}{r^{2 n-2}}+\frac{r^{2}}{l^{2}},
\ee
where $k=1,0,-1$ respectively describe the spherical case, flat case and hyperbolic case, if no special instructions are given, we mainly focus on the case $k=-1$ in the following. $m$ and $\tilde{q}$ are two integration constants, $d \Sigma_{n}^{2}$ stands for the line element for an n-dimensional hypersurface with
negative constant curvature $-n(n-1)$. The authors of \cite{7} found that the hyperbolic black hole is supersymmetric in the case of $m=0$ and $\tilde{q}=\frac{l^2}{4}$, the metric can be expressed as $f=\left(\frac{l}{2 r}-\frac{r}{l}\right)^{2}$, which is an extremal black hole solution with vanishing Hawking temperature and the mass parameter is zero. When $m=0$, there are still two black hole horizons $r_{1,2}^{2}=l^{2}\left(1 \pm \sqrt{1-4 \tilde{q}^{2} / l^{2}}\right) / 2$  provided  $4 \tilde{q}^{2} / l^{2}<1$ in solution (\ref{metric}) in the four-dimensional case. In fact, even the mass parameter $m$ is negative down to
\be
m_{c}=-2 r_{c}^{n-1}\left(1-\frac{n r_{c}^{2}}{(n-1) l^{2}}\right)
\ee
with
\be
\frac{r_{c}^{2}}{l^{2}}=\frac{n-1}{n+1}\left(1+\frac{\tilde{q}^{2}}{r_{c}^{2 n-2}}\right)
\ee
As mentioned above, the hyperbolic black hole still has a black hole structure when $\tilde{q}<0$. In this paper, we restricted ourselves to the case $\tilde{q}>0$. For the 4-dimensional hyperbolic black hole (\ref{metric}), the mass $M$, Hawking temperature $T$ , and entropy $S$ are given by\cite{Cai:2004pz}
\be
M=\frac{r_{+} \Omega\left(\frac{r_{+}^{2}}{l^{2}}+\frac{\tilde{q}^{2}}{r_{+}^{2}}-1\right)}{8 \pi G},
\ee
\be\label{TTT}
T=\frac{\frac{3 r_{+}^{2}}{l^{2}}-\frac{\tilde{q}^{2}}{r_{+}^{2}}-1}{4 \pi r_{+}},
\ee
\be\label{SSS}
S=\frac{r_{+}^{2} \Omega}{4}.
\ee
where $\Omega$ is the volume of the unit hyperbolic surface. We can easily get
\begin{equation}
C_{V}=T\left(\frac{\partial S}{\partial T}\right)_{V}=0.
\end{equation}
The specific heat capacity of a constant pressure $C_{p}$ can be calculated at the horizon $r = r_{+}$
\begin{equation}
C_{p}=\frac{r_{+}^{2} \Omega\left(3 r_{+}^{4}-l^{2}\left(\tilde{q}^{2}+r_{+}^{2}\right)\right)}{2 l^{2}\left(3 \tilde{q}^{2}+r_{+}^{2}\right)+6 r_{+}^{4}},
\ee
where we set $G=1$. Next we will build a holographic heat engine cycle similar to Fig.\ref{P-V diagram}. In the process from 1 to 2, the $Q_{H}$ is
\begin{equation}
\begin{aligned}
Q_{H}&=\int_{T_{1}}^{T_{2}} C_{P} d T=\int_{r_{1}}^{r_{2}} C_{P} \frac{dT}{d r_{+}} d r_{+} \\
&=\int_{r_{1}}^{r_{2}} \frac{6 r_{+}^{6} \Omega-3 l^{2} \Omega\left(\tilde{q}^{2}+r_{+}^{4}\right)}{8 \pi l^{2} r_{+}^{3}} d r_{+}
&=\frac{r_{+} \Omega\left(\frac{r_{+}^{2}}{l^{2}}+\frac{\tilde{q}^{2}}{r_{+}^{2}}-1\right)}{8 \pi }\bigg|^{r = r_{1}}_{r = r_{2}}=M_{2}-M_{1},
\end{aligned}
\end{equation}
Similarly, $Q_{C}$ is given by
\begin{equation}
Q_{c}=\int_{T_{4}}^{T_{3}} C_{P} d T=\int_{T_{4}}^{T_{3}}\frac{6 r_{+}^{9} \Omega-3 l^{2} r_{+}^{3} \Omega\left(\tilde{q}^{2}+r_{+}^{4}\right)}{4 l^{2}\left(5 \tilde{q}^{2}+r_{+}^{4}\right)+8 r_{+}^{6}} d T=M_{3}-M_{4}.
\end{equation}
The efficiency is
\begin{equation}
\eta=1-\frac{Q_{C}}{Q_{H}}=1-\frac{M_{3}-M_{4}}{M_{2}-M_{1}}.
\end{equation}

We can see that this efficiency is the same as the result obtained by Johnson in \cite{Johnson:2016pfa}. Next, we will compare the efficiency of the hyperbolic black hole with the well-known Carnot efficiency. We rewrite the temperature (\ref{TTT}) and the entropy (\ref{SSS}) as
\be
l_{1}=l_{2}=\frac{\sqrt{2} r_{2}^{3}}{\sqrt{\tilde{q}^{2}+2 \pi r_{2}^{5} T_{2}+r_{2}^{4}}},
\ee
\be
l_{3}=l_{4}=\frac{\sqrt{2} r_{4}^{3}}{\sqrt{\tilde{q}^{2}+2 \pi r_{4}^{5} T_{4}+r_{4}^{4}}},
\ee
\be
r_{1}=r_{4}=\frac{2^{2 / 3} \sqrt[3]{S_{4}}}{\sqrt[3]{\Omega}},
\ee
\be
r_{3}=r_{2}=\frac{2^{2 / 3} \sqrt[3]{S_{2}}}{\sqrt[3]{\Omega}}.
\ee
\begin{figure}
\centering
\includegraphics[width=6cm]{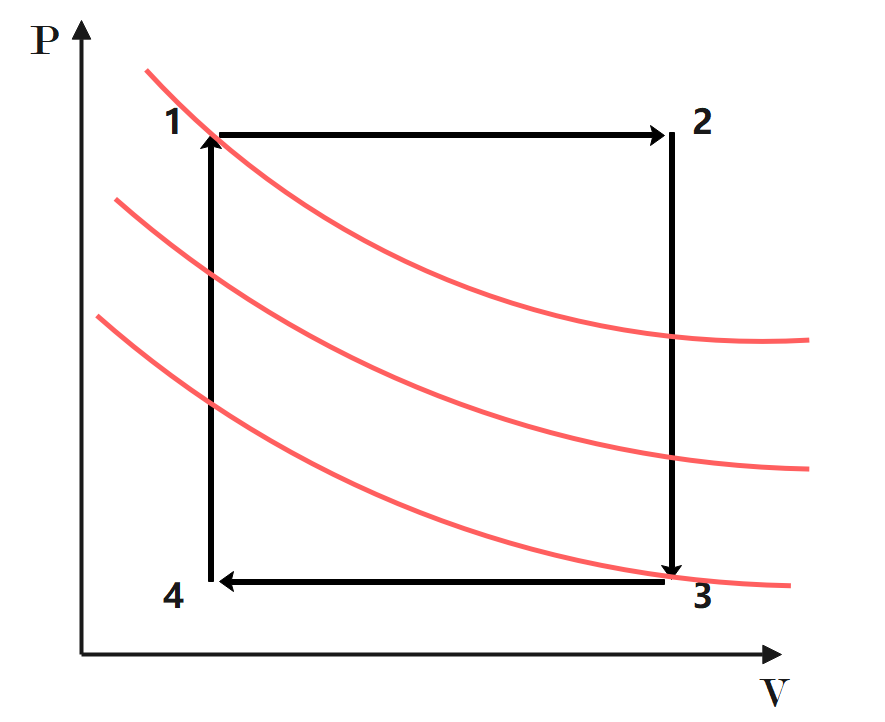}
\caption{Isotherms and rectangular heat engine cycle}
\label{P-V2 diagram}
\end{figure}
The efficiency can be rewritten as
\begin{equation}\label{Efficiency1}
\begin{aligned}
\eta &=1-\frac{M_{3}-M_{4}}{M_{2}-M_{1}}=1-\frac{\frac{3 r_{3}^{2} \Omega\left(\frac{r_{3}^{2}}{l_{3}^{2}}+\frac{\tilde{q}^{2}}{r_{3}^{4}}-1\right)}{16 \pi }-\frac{3 r_{4}^{2} \Omega\left(\frac{r_{4}^{2}}{l_{4}^{2}}+\frac{\tilde{q}^{2}}{r_{4}^{4}}-1\right)}{16 \pi }}{\frac{3 r_{2}^{2} \Omega\left(\frac{r_{2}^{2}}{l_{2}^{2}}+\frac{\tilde{q}^{2}}{r_{2}^{4}}-1\right)}{16 \pi }-\frac{3 r_{1}^{2} \Omega\left(\frac{r_{1}^{2}}{l_{1}^{2}}+\frac{\tilde{q}^{2}}{r_{1}^{4}}-1\right)}{16 \pi }} \\
&=\frac{\left(S_{2}^{2 / 3}+S_{4}^{2 / 3}\right)A-\sqrt[3]{2} \tilde{q}^{2} S_{4}^{2} \Omega^{5 / 3}-16 \times 2^{2 / 3} \pi S_{2}^{5 / 3} S_{4}^{2} T_{2}-8 S_{2}^{4 / 3} S_{4}^{2} \sqrt[3]{\Omega}}{\sqrt[3]{2} \tilde{q}^{2} S_{2}^{2 / 3} S_{4}^{2} \Omega^{5 / 3}+S_{2}^{4 / 3}\left(8 S_{4}^{8 / 3} \sqrt[3]{\Omega}-2 \sqrt[3]{2} \tilde{q}^{2} S_{4}^{4 / 3} \Omega^{5 / 3}\right)+B}.
\end{aligned}
\end{equation}
where
\be
\begin{aligned}
A&=\sqrt[3]{2} \tilde{q}^{2} S_{2}^{2} \Omega^{5 / 3}+16 \times 2^{2 / 3} \pi S_{2}^{2} S_{4}^{5 / 3} T_{4}+8 S_{2}^{2} S_{4}^{4 / 3} \sqrt[3]{\Omega}, \\
B&=\sqrt[3]{2} \tilde{q}^{2} S_{4}^{8 / 3} \Omega^{5 / 3}+16 \times 2^{2 / 3} \pi S_{2}^{7 / 3} S_{4}^{2} T_{2}-8 S_{2}^{2} S_{4}^{2} \sqrt[3]{\Omega}+16 \times 2^{2 / 3} \pi S_{2}^{5 / 3} S_{4}^{8 / 3} T_{2}.
\end{aligned}
\ee
\begin{figure}[h]
\centering
\includegraphics[width=6cm]{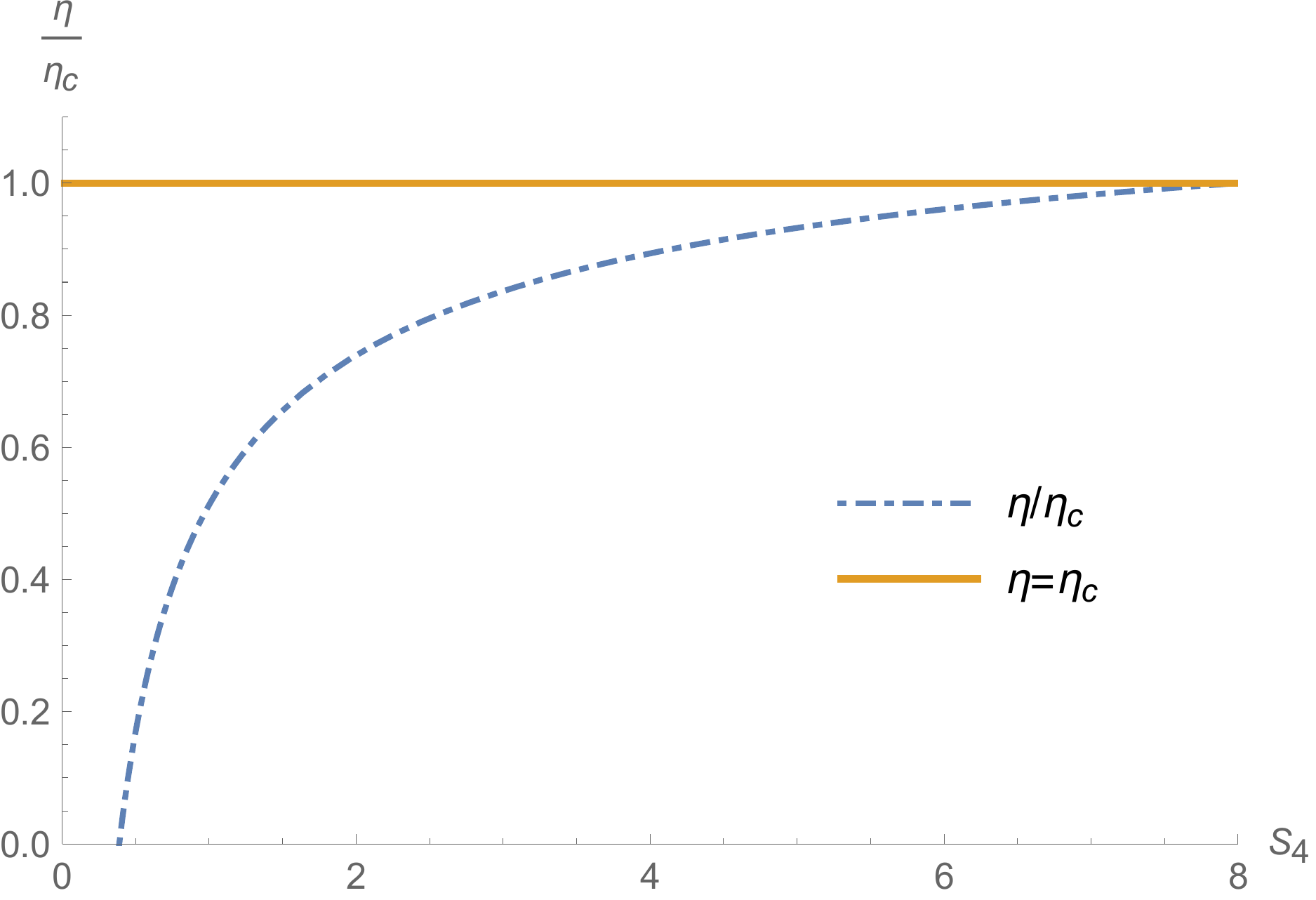}
\caption{The vertical axis in the figure represents the ratio of the hyperbolic black hole rectangular cycle efficiency $\eta$ to the Carnot efficiency $\eta_{C}$, and the horizontal axis is the entropy $S_{4}$. The blue line represents the heat engine efficiency of the 4-dimensional hyperbolic black hole when $\tilde{q}=1$. We set $\Omega=1, T_{2}=5, T_{4}=1, S_{2}=8$.}
\label{ncnS4 diagram}
\end{figure}
Considering the isotherm shown in Fig.\ref{P-V2 diagram}, we can find that $T_{2}$ is the highest temperature ($T_{2}=T_{H}$) and $T_{4}$ is the lowest temperature ($T_{4}=T_{C}$). Note that when $S_{4} \rightarrow S_{2}$, the efficiency (\ref{Efficiency1}) can be written as
\begin{equation}
\eta_{C}=1-\frac{T_{C}}{T_{H}},
\end{equation}
\begin{figure}[h]
    \centering    	    	                   	
    \includegraphics[scale=0.4]{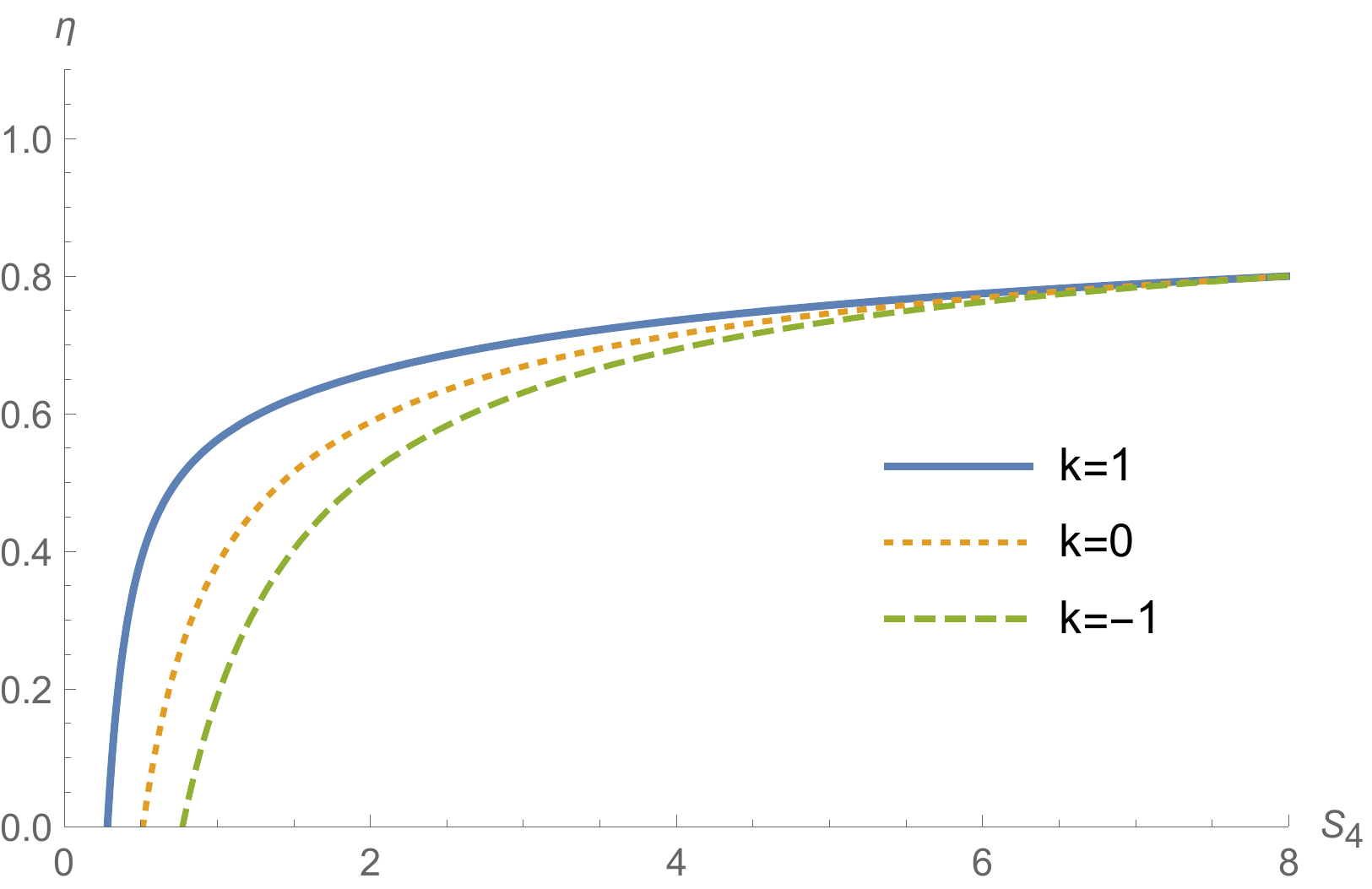}
    \includegraphics[scale=0.4]{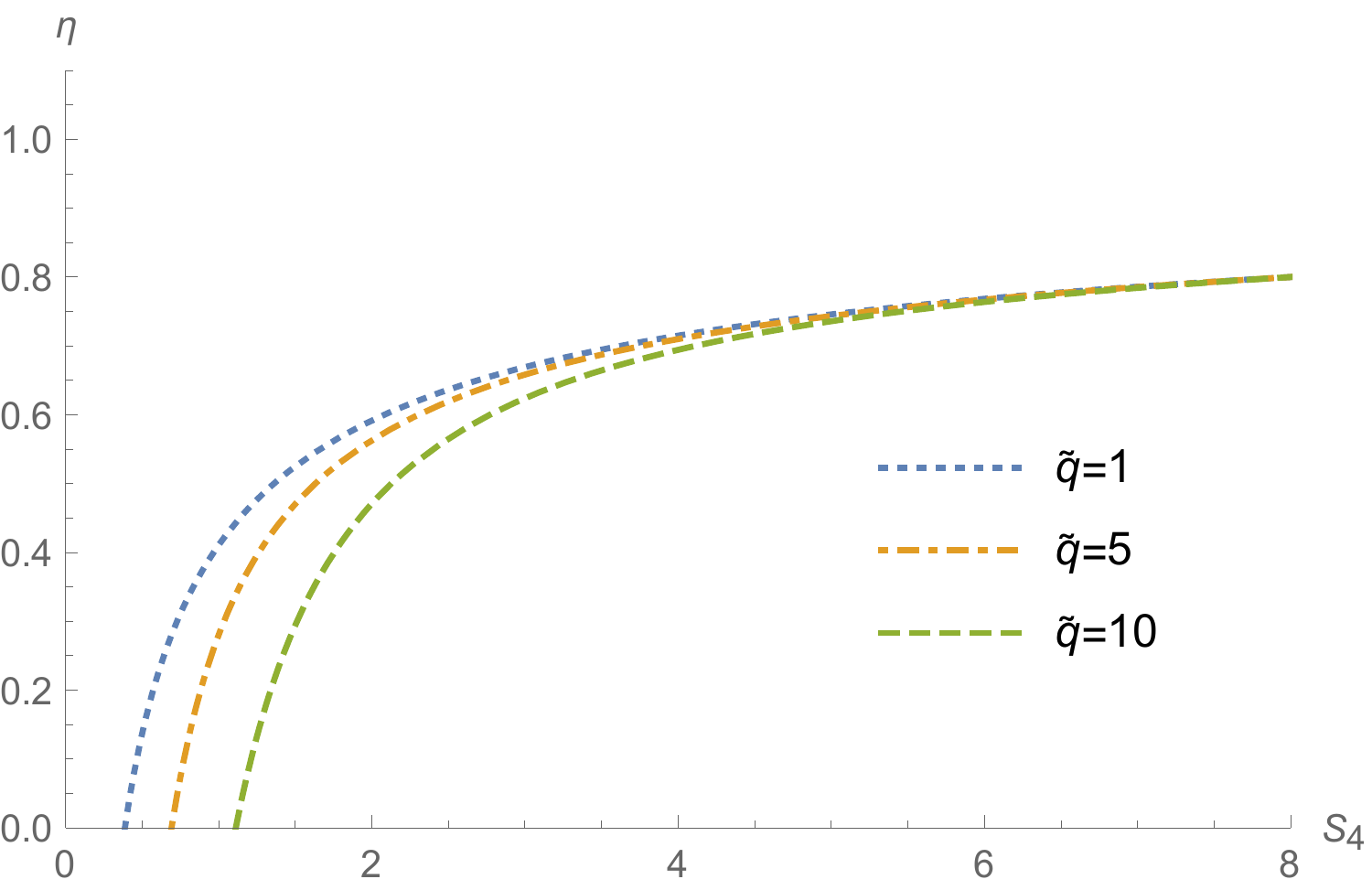}
    \caption{Left figure: black hole heat engine efficiency with different $k$, where we choose $\tilde{q}=0.1$. Right figure: the efficiency with different electric charges $\tilde{q}=1,5,10$.}
    \label{101 diagram}
\end{figure}

which is the efficiency of the Carnot cycle. In other words, when the entropy (volume) change is small enough ($\Delta S=S_{2}-S_{1}=S_{3}-S_{4}$), the rectangular cycle becomes the Carnot cycle (As shown in Fig.\ref{ncnS4 diagram}).

Furthermore, we can obtain the efficiencies of spherical ($k=1$) and flat ($k=0$) case in the same way. The results (as shown in the left picture of Fig.\ref{101 diagram}) show that the efficiency of hyperbolic black holes are lower than the cases of $k=1$ and $k=0$. Note that we restricted ourselves to the case $\tilde{q}>0$, Since when $\tilde{q}=0$, the black hole with $k=0,-1$ is not physical. In addition, the right picture of Fig.\ref{101 diagram} shows the hyperbolic black hole effect with different electric charges. The results show that less charge corresponds to higher efficiency in the case of $\tilde{q}>0$, because the charge confines part of the energy.


\section{Benchmark cycle for hyperbolic charged black hole}
In fact, since the rectangles can be added, we can use equation (\ref{exact efficiency}) as a basis to calculate any closed loop on the P-V graph\cite{Johnson:2016pfa} (as shown in Fig.\ref{Add rectangles}). Note that only the upper and lower edges of the rectangle will generate heat flow.
\begin{figure}[h]
\centering
\includegraphics[width=6cm]{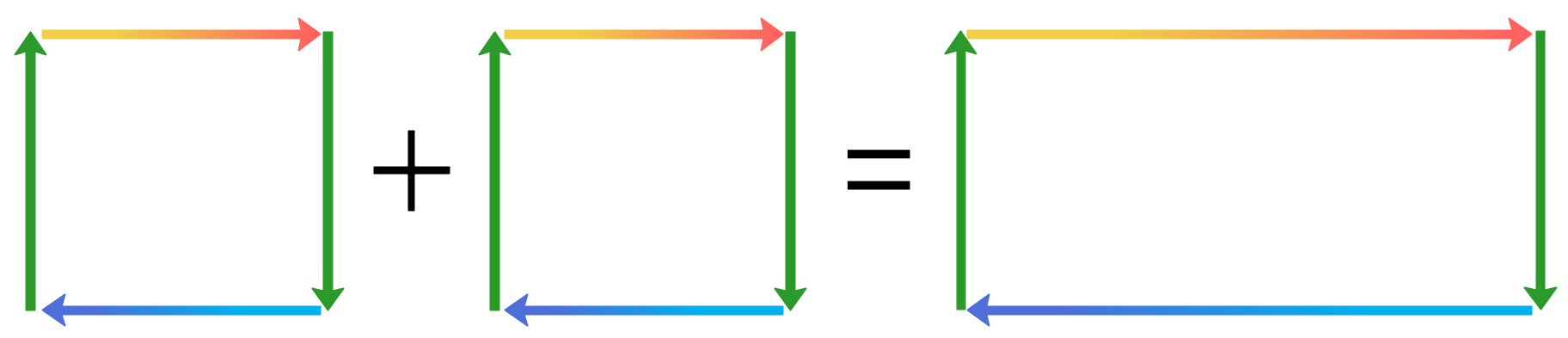}
\caption{Adding cycles that share an edge.}
\label{Add rectangles}
\end{figure}
In this section, we introduce a benchmark cycle \cite{Chakraborty:2016ssb,Chakraborty:2017weq}, which could be parameterised as a circle. We cover a circle with a center of $(P_{0}, V_{0})$ and radius $R$ on $N*N$ regular lattice of squares, where $N$ is an even number.
 $Q_{H}$ is determined by the sum of the mass differences at both ends of each ``hot cell'' isobar. The cell refers to the small square in Fig.\ref{Add rectangles}. Edge cells are called hot cells if they have their upper edges open, and cold cells if they have their lower edges open. $Q_{C}$ is determined by ``cold cell''.
\be
\eta=1-\frac{Q_{C}}{Q_{H}}, \quad Q_{H}=\sum_{ \text { ith hot cell }}\left(M_{2}^{(i)}-M_{1}^{(i)}\right), \quad Q_{C}=\sum_{\text {ith cold cell }}\left(M_{3}^{(i)}-M_{4}^{(i)}\right).
\ee

\begin{figure}[h]
    \centering    	    	                   	
    \includegraphics[scale=0.4]{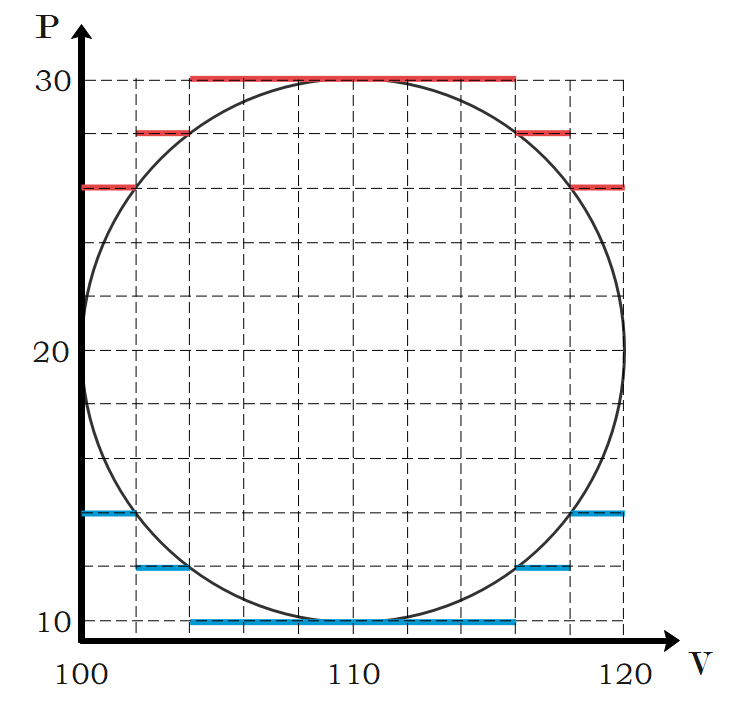}
    \includegraphics[scale=0.4]{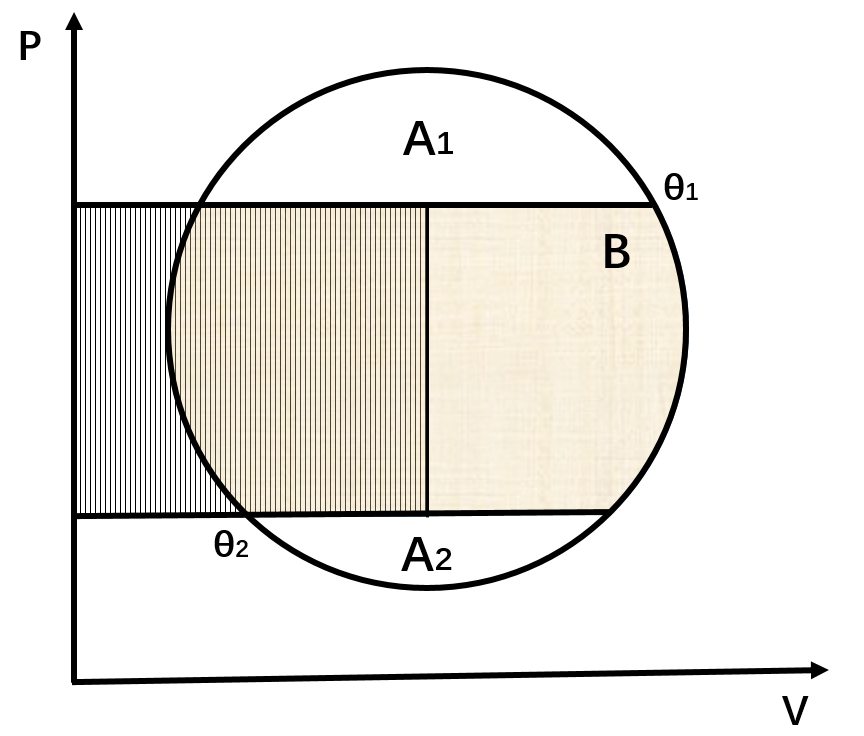}
    \caption{Left: Example of covering a circle on 100 regular lattice of squares ($N=10$). The origin is at $(P_{0}, V_{0})=(20, 110)$ and the radius is $R=10$. The red line represents the top edge of the hot cell, and the blue line represents the bottom edge of the cold cell. As $N$ increases, these lines converge to the boundary of the circle. Right: A general benchmarking cycle, partitioned into subregions.}
    \label{benchmark C}
\end{figure}

A more general expression of efficiency can be obtained through the first law of thermodynamics. Consider for convenience a circular benchmarking cycle defined parametrically by
\begin{equation}
\begin{array}
{l}V(\theta)=V_{0}+R \cos \theta \\P(\theta)=P_{0}-R \sin \theta .
\end{array}
\end{equation}
The first law of thermodynamics in the canonical ensemble can be expressed as
\begin{equation}
\delta Q=T d S=d M-V d P \Rightarrow Q=\Delta M-\int V d P
\end{equation}
The first law can help us to express the integral of $Q_{H}$ and $Q_{C}$ in the process by the area of the circle and the difference in mass
 (see Fig\ref{benchmark C}). The area of the circle is divided into two parts by $Q_{1}$ and $Q_{2}$, and two lines parallel to the $V$ axis divide the circle into three parts, $A_{1}$, $A_{2}$, and $B$ ($B =\pi R^{2}-A_{1}-A_{2}$). Then we can get the $Q_{H}$ and $Q_{C}$\cite{33}
\begin{equation}
 \begin{aligned}
 Q_{H} &=M_{1}-M_{2}-V_{0}\left(P_{1}-P_{2}\right)+\frac{1}{2} B+A_{1} \\
-Q_{C} &=M_{2}-M_{1}+V_{0}\left(P_{1}-P_{2}\right)+\frac{1}{2} B+A_{2}
 \end{aligned}
\end{equation}
Efficiency can be expressed as
\begin{equation}
\eta=\frac{\pi R^{2}}{M_{1}-M_{2}-V_{0} R\left(\sin \theta_{2}-\sin \theta_{1}\right)+\frac{R^{2}}{2}\left[\theta_{1}-\theta_{2}+\sin \left(\theta_{1}-\theta_{2}\right) \cos \left(\theta_{1}-\theta_{2}\right)\right]}
\end{equation}

In particular, when the specific heat capacity at constant volume vanished ($C_{V} = 0$), the above equation can be rewritten as
\cite{Hennigar:2017apu}
\be
\eta=\frac{\pi R^{2}}{\pi R^{2} / 2+\Delta M},
\ee
where $\Delta M$ is the enthalpy difference of two points at the left and right ends of the circle
\be
\Delta M=M\left(V_{o}+R, P_{o}\right)-M\left(V_{0}-R, P_{0}\right).
\ee
Based on the thermodynamic quantities of the charged hyperbolic black hole, we have obtained the four-dimensional heat engine efficiency by numerical simulation. We choose the circle whose origin is at $(P_{0}, V_{0})=(20, 110)$ and the radius is $R=10$ (shown in right Fig.\ref{benchmark C}).
\begin{figure}
\centering
\includegraphics[width=6.5cm]{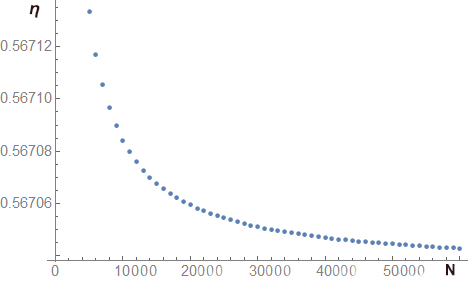}
\caption{The efficiency of benchmarking cycle as a function of regular lattice of squares N. For $N = 50000$,  $\eta$ converge to $0.56704$. Here the origin is $(P_{0}, V_{0})=(20, 110)$ and the radius is $R=10$, we take $\tilde{q}=0.5 , G=1$.}
\label{HN}
\end{figure}
The efficiency of charged hyperbolic black holes (as shown in Fig.\ref{HN}) converges to $0.56704$.
In fact, we can also rewrite temperature as a function of pressure and volume $T(P,V)$, the efficiency of the Carnot cycle is $\eta_{C}=0.66677$. We can see that the efficiency of hyperbolic black holes $\eta=0.56704$ is lower than the Carnot efficiency $\eta_{C}=0.66677$.

\section{Conclusion and discussion}
We consider a four-dimensional charged hyperbolic black hole as a working matter to establish a black hole holographic heat engine, and use the rectangular cycle to obtain the heat engine efficiency. The efficiency is consistent with the results in the original literature\cite{Johnson:2016pfa}. Furthermore, we compare the result with the Carnot efficiency and find that when the change in entropy vanished, the charged hyperbolic black hole heat engine becomes a reversible heat engine whose efficiency is the Carnot efficiency.
In addition, we also study the efficiency of the flat case and the spherical case. The results show that the efficiency in the spherical case is higher than the flat case and the hyperbolic case. We also found that less charge corresponds to higher efficiency in the case of $\tilde{q}>0$. At last, when calculating in benchmark scheme, we find that the efficiency of charged hyperbolic black holes is always lower than the Carnot efficiency.


\end{CJK*}  
\end{document}